\def\Mol{M{\o}ller\ }
\def\ssw{s_W^2}
\def\sswq{s_W^4}

\def\sll{\sigma_{{\rm LL}}}
\def\slr{\sigma_{{\rm LR}}}
\def\srl{\sigma_{{\rm RL}}}
\def\srr{\sigma_{{\rm RR}}}

\def\dsll{{\rm d}\sigma_{{\rm LL}}}
\def\dslr{{\rm d}\sigma_{{\rm LR}}}
\def\dsrl{{\rm d}\sigma_{{\rm RL}}}
\def\dsrr{{\rm d}\sigma_{{\rm RR}}}

\def\nll{N_{{\rm LL}}}
\def\nlr{N_{{\rm LR}}}
\def\nrl{N_{{\rm RL}}}
\def\nrr{N_{{\rm RR}}}

\def\peff{P_{{\rm eff}}}
\def\pe{P_{e^-}}
\def\pp{P_{e^+}}
\def\msbar{{\overline{\rm MS}}}
\def\alr{A_{\rm LR}}
\def\ep{$e^+e^-$ }
\def\mol{$e^-e^-$ }
\def\cw{\cos^2\theta_W}
\def\sw{\sin^2\theta_W}
\def\swq{\sin^4\theta_W}

\def\ee{\end{eqnarray}}
\def\be{\begin{eqnarray}}
\newcommand{\gsim}{\lower.7ex\hbox{$
\;\stackrel{\textstyle>}{\sim}\;$}}
\newcommand{\lsim}{\lower.7ex\hbox{$
\;\stackrel{\textstyle<}{\sim}\;$}}

\documentstyle[ee,twoside,psfig]{article}

\begin{document}
\normalsize\textlineskip

\begin{flushright}
BNL-HET-98/4 \\
hep-ph/9801394 \\ January 1998
\end{flushright}

\title{PARITY VIOLATING ASYMMETRIES\\[2mm] AT FUTURE LEPTON COLLIDERS}
\author{ANDRZEJ CZARNECKI and WILLIAM J. MARCIANO} 
\address{
Brookhaven National Laboratory\\  Upton, New York 11973}

\maketitle\abstracts{
Parity violating left--right scattering asymmetries at future high
energy \ep and \mol colliders are examined. The utility of two
polarized beams for precision measurements is pointed out. Sensitivity
to ``new'' short--distance physics is briefly discussed. Electroweak
radiative corrections due to the running of $\sw(q^2)$ are exhibited.
}

\vspace*{5mm}

The International Linear Collider (ILC) will have tremendous discovery
poten\-tial\cite{Kuhl}. Combining high energy and high luminosity
capabilities, it will thoroughly explore \ep col\-li\-sions from
$\sqrt{s}\approx 500$ GeV to 1.5 TeV. In addition, the possibility
exists to run at lower energies, such as the $Z$ pole $\sim 91$ GeV,
while retaining relatively high luminosity. There, one could carry out
very precise measurements and perform detailed studies of $b$ and
$\tau$ physics. Indeed, one could imagine repeating the extremely
successful LEP I program, but with more than 100 fold increase in
luminosity and with a highly polarized $e^-$ beam ($P_{e^-}\sim
90$\%). It might even be possible to employ polarized 
positrons\cite{Kuhl}. Other complementary ILC options include
$e^-e^-$, $\gamma\gamma$, and  $e^-\gamma$
collider programs, also utilizing high luminosity and polarization.

The \mol collider offers some particularly exciting
features\cite{Heusch}. For example, it can presumably employ two
polarized $e^-$ beams with $P_1$ and $P_2\approx$ 90\% while
maintaining high luminosity. The polarization can be a powerful tool
for suppressing backgrounds as well as for sorting out properties of
``new physics'' such as supersymmetry, $Z'$ bosons, strong dynamics,
compositeness etc.  It can also be useful for carrying out precision
electroweak measurements, as we shall see.

In this paper, we examine parity violating \ep and \mol scattering
asymmetries at future lepton colliders using single or double beam
polarizations.  The power of polarization has already been
demonstrated at the SLC where $\sw$ has been very precisely determined
using $e^-$ polarization ($P_{e^-}\approx 77$\%) at the $Z$ 
pole\cite{Abe804}.  The left--right asymmetry for \ep$\to$ hadrons
\be
\alr \equiv 
{\sigma(e^+e^-_L\to \mbox{hadrons}) - \sigma(e^+e^-_R\to \mbox{hadrons}) 
\over
\sigma(e^+e^-_L\to \mbox{hadrons}) + \sigma(e^+e^-_R\to \mbox{hadrons}) 
}
\ee
has been measured to about $\pm 2$\% and used to obtain
$\sw(m_Z)_\msbar$ to about $\pm 0.0004$.  Ongoing efforts are expected
to further reduce those errors by about a factor of 2 and thus provide
the most accurate measurement of that fundamental parameter.

At the $Z$ pole, $\alr$ is predicted (at tree level) to be
\be
\alr = {2(1-4\sw)\over 1+(1-4\sw)^2}.
\ee
Because $\sw(m_Z)_\msbar \approx 0.23$ is near $1/4$, where $\alr$
vanishes and is sensitive to small changes in $\sw$, it can be
extracted with high precision.  Indeed, $\Delta\sw/\sw\approx {1\over
10} \Delta \alr/\alr$. Furthermore, an asymmetry ratio has small
systematic uncertainties. At the SLC, the dominant systematic error is
a $\pm 0.5$\% polarization uncertainty which contributes a $\pm
0.0001$ uncertainty in $\sw$.

Because the $e^-$ polarization is not 100\%, the $\alr$ experiment
actually measures (after acceptance and other small corrections)
\be
{N_L-N_R\over N_L+N_R} = P_{e^-}\alr,
\ee
where $N_i$, $i=L,R$, are the number of events for each polarization
setting.
Hence, the $\pm 0.5$\% uncertainty in $P_{e^-}$ leads to about a $\pm
0.05$\% systematic error in $\sw$. If both beams were polarized, one
would measure
\be
\lefteqn{\hspace*{-7mm}{N_{LR}-N_{RL} \over  N_{LR}+N_{RL}} = \peff \alr,}
\nonumber \\
&&\hspace*{-15mm}\peff = {\pe -\pp\over 1-\pe\pp},
\label{eq:peff}
\ee
where it is assumed that $\pe$ and $\pp$ have opposite signs and can
each be individually alternated in sign from pulse to pulse. The
effective polarization, $\peff$, is larger than either $\pe$ or $\pp$
and, as we shall subsequently show, has a relatively small
uncertainty.

One could envision a dedicated run at the $Z$ pole after the ILC is
completed. With $\gsim 10^8$ $Z$ bosons (relatively easy to obtain at
ILC luminosities) and $\pe \approx 90$\%, $\alr$ could be
statistically determined to better than $\pm 0.1$\% and $\sw$ to $\pm
0.01$\%, i.e. $\pm0.00002$. Such an incredibly precise measurement of
the weak mixing angle is highly desirable. Taken together with
$G_\mu$, $m_Z$, $\alpha$, and $m_t$ it could be used to predict the
Higgs scalar mass\cite{DGP188} to about $\pm 5$\% or constrain ``new
physics'' loop effects. Of course, to fully utilize such precision,
hadronic loop uncertainties, two loop electroweak corrections and
$\Delta m_t$ would also need to be improved.  In addition, the
polarization uncertainty and other systematics would have to be
controlled to better than $\pm 0.1$\%, a challenging requirement.
Currently $\Delta
\pe/\pe$ is about $\pm 0.5$\% at the SLC\cite{Abe804}.

If both $e^-$ and $e^+$ beams could be polarized, the overall
polarization uncertainty in $\alr$ would be naturally reduced. In that
case the relevant quantity is $\peff$ (see Eq.~(\ref{eq:peff})). One
finds
\be
{\Delta\peff\over \peff} = 
{(1-\pp^2)\pe \over (\pe - \pp)(1-\pe\pp)}
{\Delta \pe\over \pe}
-
{(1-\pe^2)\pp \over (\pe - \pp)(1-\pe\pp)}
{\Delta \pp\over \pp}.
\ee
So, for example, if $|\pe|=0.9000\pm 0.0045$ and  $|\pp|=0.6500\pm
0.0065$ (i.e. $\pm 1$\% for the $e^+$ beam), the effective
polarization would be
\be
\peff=0.9779\pm 0.0012,
\ee
very near $\pm 0.1$\% uncertainty. With both beams polarized, one
could even directly measure the polarizations to very high
precision. Indeed, by alternating the polarization of each beam on a
pulse by pulse basis, one would measure the number of events for each
of the four polarization configurations $\nll$, $\nlr$, $\nrl$, and
$\nrr$. Because the beams are not fully polarized, $\nll $ and $\nrr$
do not vanish. In fact, the ratio of the expected rates (up to the
left--right asymmetry) should be
\be
\lefteqn{\nll:\nlr:\nrl:\nrr }
\nonumber \\
&&
::
(1-|\pe\pp|):(1+|\pe\pp|):(1+|\pe\pp|):(1-|\pe\pp|).
\ee
So, for $\pe \approx 0.9$ and $\pp=0.65$, the $\nll$ and $\nrr$ are
only about a factor of 4 smaller than $\nlr$ and $\nrl$.   One can then
determine $\pe$ and $\pp$ via
\be
{\nll+\nlr-\nrl-\nrr \over
\nll+\nlr+\nrl+\nrr }
&=&\phantom{-} \pe \alr,
\nonumber \\
{\nrr+\nlr-\nrl-\nll \over
\nrr+\nlr+\nrl+\nll }
&=& -\pp \alr,
\label{eq:ens}
\ee
used in conjunction with Eq.~(\ref{eq:peff}).  Hence, with high
luminosity and both beams polarized one can use the $N_{ij}$ to
simultaneously determine $\pe $, $\pp$ and $\alr$ via
Eqs.~(\ref{eq:peff}) and (\ref{eq:ens}) to about $\pm$0.1\%.

If polarized positrons are not possible, one might hope to carry out
the above studies at a high luminosity $\mu^+\mu^-$ collider with both
beams polarized. There, the spectrum of the decaying muons would also
provide a very precise direct polarization measurement. 

In the case of higher energy \ep collisions, Cuypers and
Gambino\cite{Cuypers:1996it} have examined Bhabha scattering
$e^+e^-\to e^+e^-$ using polarized beams and concluded
$\Delta\sw=\pm0.0004$ at $\sqrt{s}=500 $ GeV and $\Delta\sw=\pm0.0001$
at $\sqrt{s}=2$ TeV were possible for $|\pp|=|\pe|=0.9$. Although not
quite competitive with a potential future $Z$ pole measurement, such
studies would provide powerful constraints on ``new physics,'' or
signals of its presence. For example, Barklow\cite{BarklowH} has shown
that electron compositeness parametrized by an effective interaction
\be
{2\pi\over \Lambda^2}
 \overline{e}_L \gamma_\mu e_L
 \overline{e}_L \gamma^\mu e_L
\label{eq:fourf}
\ee
can be probed up to $\Lambda\approx 100$ TeV for $\sqrt{s}=1 $ TeV,
$\pe=90$\%, and $\pp=0$.

One could also study \ep$\to \overline{f}f$ ($f\ne e$) with polarized
beams above the $Z$ pole. At tree level, the standard model predicts
(with $s_W\equiv \sin\theta_W$)
\be
\lefteqn{\hspace*{-6mm} \alr(e^+e^- \to \mu^+\mu^- \mbox{ or } \tau^+\tau^-)
=(1-4\ssw)F(s/m_Z^2)}
\nonumber\\[3mm]
F(x)&=&
{x^2(1+4\ssw)+8x\ssw(\ssw-1)\over
x^2(1+24 \sswq  )-8x(1-\ssw)(8\sswq+\ssw)+64\sswq(1-\ssw)^2}
\nonumber \\[3mm]
\lefteqn{\hspace*{-6mm}  F(x)\to {4\over 5}+{\cal O}(1-4\ssw) 
\quad (\mbox{for large  }x).}
\label{eq:10}
\ee
Similar (larger) asymmetries can be derived for hadronic
cross--sections.  (They are not suppressed by $(1-4\sw)$ above the $Z$
pole\cite{KMP}.)

Statistically, one expects to measure all such $s$--channel asymmetries
 to $\Delta\alr
\approx 0.005$ (systematics should not be a problem at that level).
Again, the main utility of such studies would not be to measure $\sw$,
but to search for or constrain ``new'' short--distance physics by
comparing experiment with the standard model prediction.  For example,
a four--fermion interaction of the form (cf.~Eq.~(\ref{eq:fourf}))
\be
{2\pi\over \Lambda^2}
 \overline{e}_L \gamma_\nu e_L
 \overline{\mu}_L \gamma^\nu \mu_L
\label{eq:fourfmu}
\ee
or a different chiral combination, would shift the asymmetry by
(roughly)
\be
\Delta\alr (e^+e^-\to \mu^+\mu^-) \sim {3\over 10}
{s\over \alpha\Lambda^2}.
\ee
At $\sqrt{s}=500 $ GeV, a $\Delta\alr\sim \pm 0.005$ would probe
$\Lambda\sim 40$ TeV while a similar sensitivity at $\sqrt{s}=1.5 $
TeV would probe $\Lambda\sim 120$ TeV. In the case of additional $Z'$
gauge bosons, such as the $Z_\chi$ of SO(10), those capabilities
translate into $m_{Z_\chi}\sim $ 3--10 TeV sensitivity.  Similar
sensitivity applies to effective four--fermion operators with
electrons and quarks.

\Mol scattering, $e^-e^-\to e^-e^-$, at the ILC can also be used for
precision tests of the standard model and searches for ``new
physics.''  Indeed, in some cases it can provide a more powerful probe
than $e^+e^-$. One can assume with some confidence that both $e^-$
beams will be polarized with $|P_1|=|P_2|=0.9$ and about $\pm 0.5$\%
uncertainty\cite{Heusch}. The effective polarization will therefore be
(with like sign $P_1$ and $P_2$)
\be
\peff = {P_1+P_2\over 1+P_1P_2} = 0.9945 \pm 0.0004
\ee
Again, we see that $\peff$ is quite large and has essentially
negligible uncertainty compared to  $P_1$ and $P_2$.  

The differential cross-section in \Mol scattering\cite{tree} is
characterized by a single parameter, the scattering angle $\theta$
with respect to the beam axis or
\be 
y={1-\cos\theta\over 2}, \qquad 0\le \theta \le \pi.
\ee
The cross--section grows as $1/y$ for small angle scattering. Hence,
high statistics are possible in the very forward region. Good angular
coverage is therefore important for precision measurements.
The variable $y$ relates $s$ and the momentum transfer $Q^2 = -q^2$
via 
\be
Q^2=ys, \qquad 0\le y\le 1.
\ee
Note, that $y$ and $1-y$ correspond to indistinguishable events. Very
forward $e^-e^-$ events will therefore be composed of high and low
$Q^2$ contributions.

One can consider two distinct but similar parity violating \Mol
asymmetries.  The single spin asymmetry is defined by\cite{tree}
\be
\alr^{(1)} \equiv 
{\dsll+\dslr-\dsrl-\dsrr \over \dsll+\dslr+\dsrl+\dsrr}
\label{eq:16}
\ee
while the double spin asymmetry is\cite{Cuypers:1996it}
\be
\alr^{(2)} \equiv 
{\dsll-\dsrr \over \dsll+\dsrr}
\label{eq:17}
\ee
where the subscripts denote the initial \mol states'
polarizations. Here, unlike $s-$channel $e^+e^-$, all the
d$\sigma_{ij}$ are non-vanishing. By rotational invariance 
d$\sigma_{LR}$ = d$\sigma_{RL}$; so, only the denominators in
Eqs.~(\ref{eq:16}) and (\ref{eq:17}) differ.

Experimentally, one can and probably will flip the individual
polarizations (pulse by pulse) and measure $\nll$, $\nlr$, $\nrl$, and
$\nrr$ (the number of events in each mode) for fixed luminosity and
polarization.  From those measurements, the polarizations and
$\alr^{(2)}(y)$ can be simultaneously determined using
\be
{\nll+\nlr-\nrl-\nrr \over\nll+\nlr+\nrl+\nrr}&=&\phantom{-}P_1\alr^{(1)}(y),
\label{eq:18}
\\
{\nrr+\nlr-\nrl-\nll \over\nrr+\nlr+\nrl+\nll}&=&-P_2\alr^{(1)}(y),
\label{eq:19}
\\
{\nll-\nrr\over \nll+\nrr} &=& \peff \alr^{(2)}(y)
\left(
{1\over 1+ {1-P_1P_2\over 1+P_1P_2}{\slr+\srl\over \sll+\srr}}
\right),
\label{eq:20}
\\
\peff&=& {P_1+P_2\over 1+P_1P_2}.
\nonumber
\ee
For $P_1=P_2=0.9$, the correction term in parentheses of
Eq.~(\ref{eq:20}) is small but must be accounted for.  Using
Eq.~(\ref{eq:20}), $\alr^{(2)}$ (which depends on $\sw$) can be
extracted from data and compared with the standard model prediction. A
deviation from expectations would signal ``new physics.''

In general the d$\sigma_{ij}$ for \Mol scattering are somewhat lengthy
expressions with contributions from direct and crossed $\gamma $ and
$Z$ exchange amplitudes (see fig.~1).
\begin{figure}[htb]
\noindent
\begin{minipage}{16.cm}
\hspace*{-5mm}
$
\mbox{
\hspace*{22mm}
\begin{tabular}{ccc}
\psfig{figure=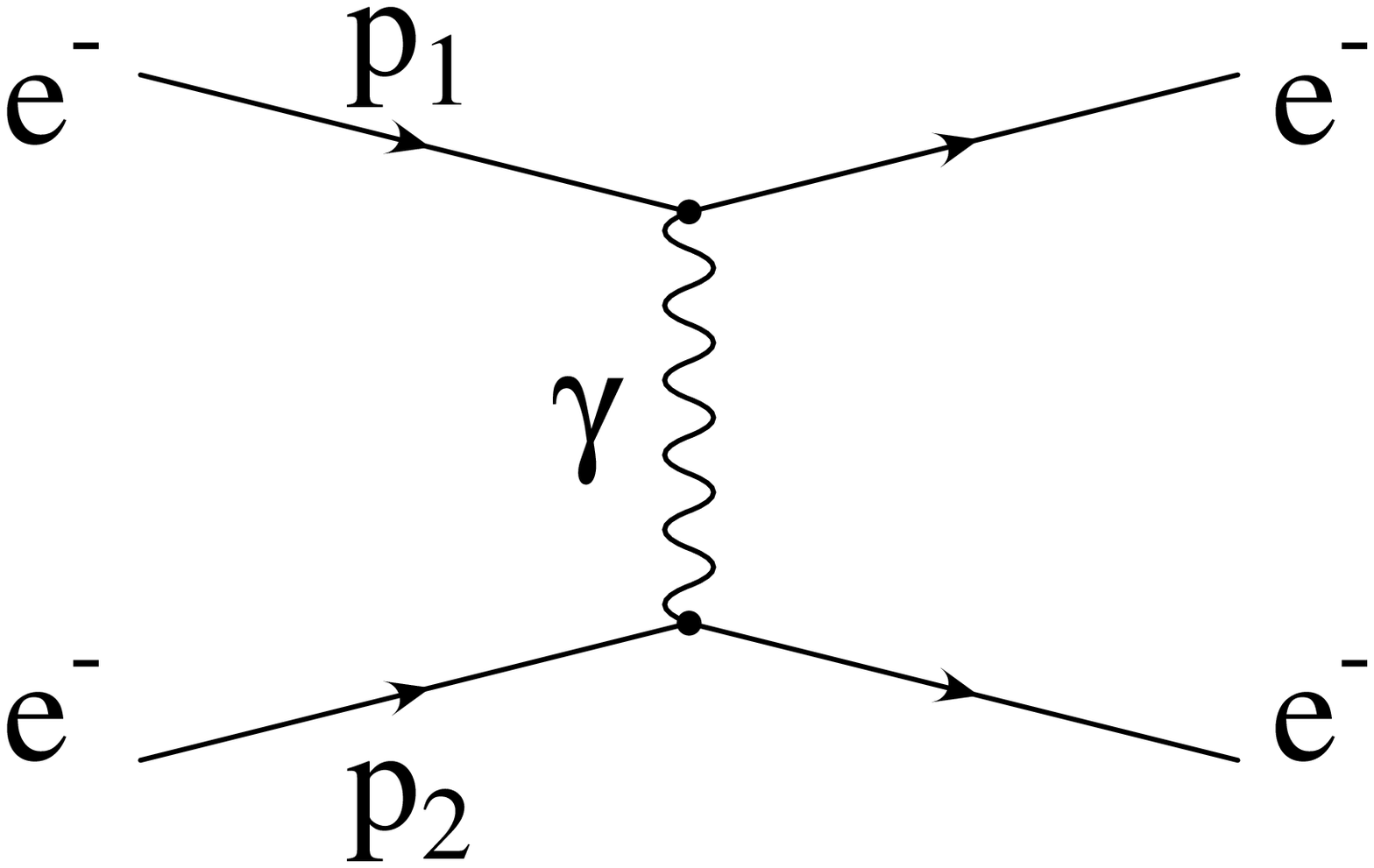,width=3.8cm,bbllx=210pt,bblly=410pt,%
bburx=630pt,bbury=550pt} 
&+&\hspace*{18mm}
\psfig{figure=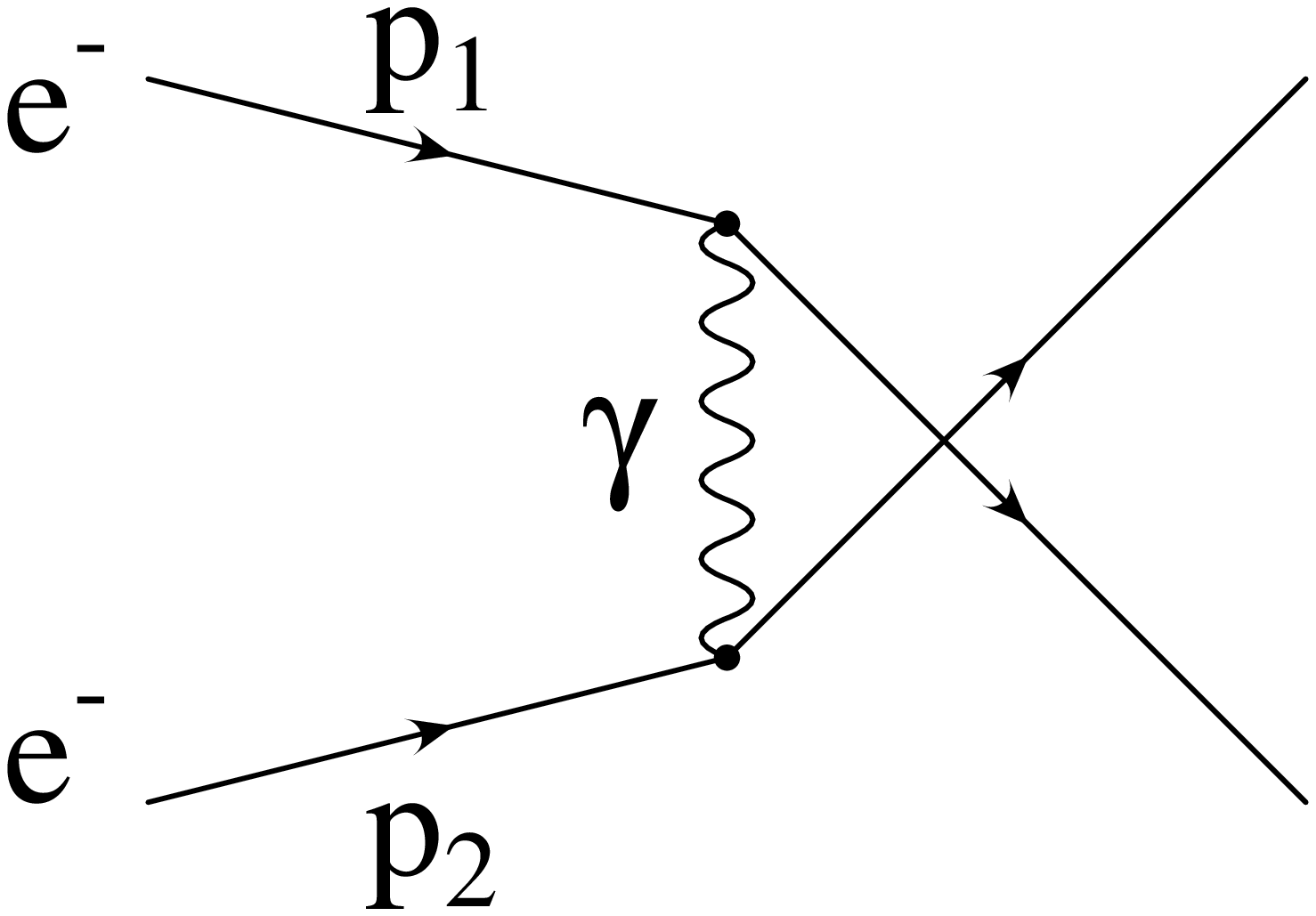,width=3.8cm,bbllx=210pt,bblly=410pt,%
bburx=630pt,bbury=550pt} \\[2cm]
\psfig{figure=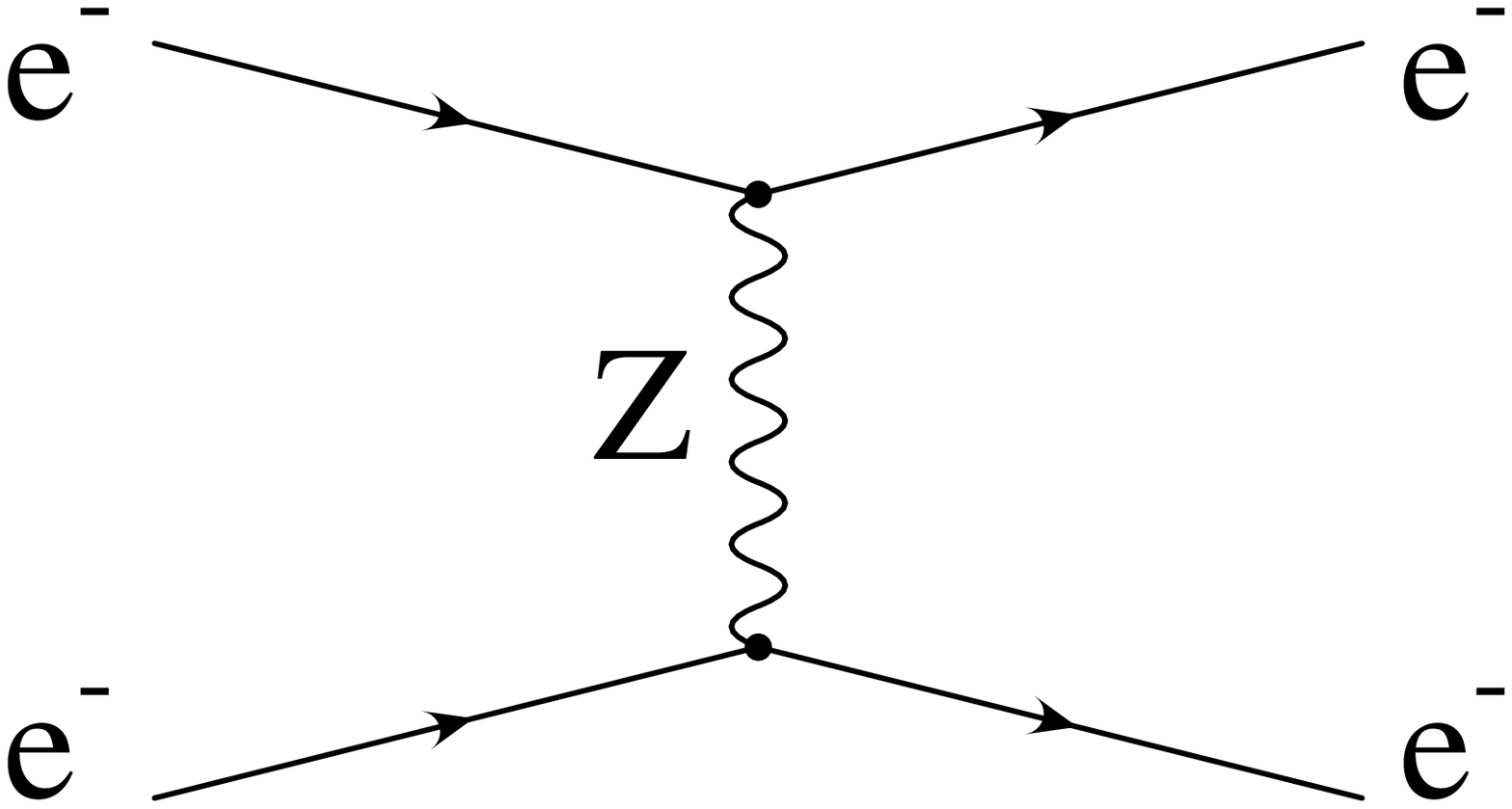,width=3.8cm,bbllx=210pt,bblly=410pt,%
bburx=630pt,bbury=550pt}
&+&\hspace*{18mm}
\psfig{figure=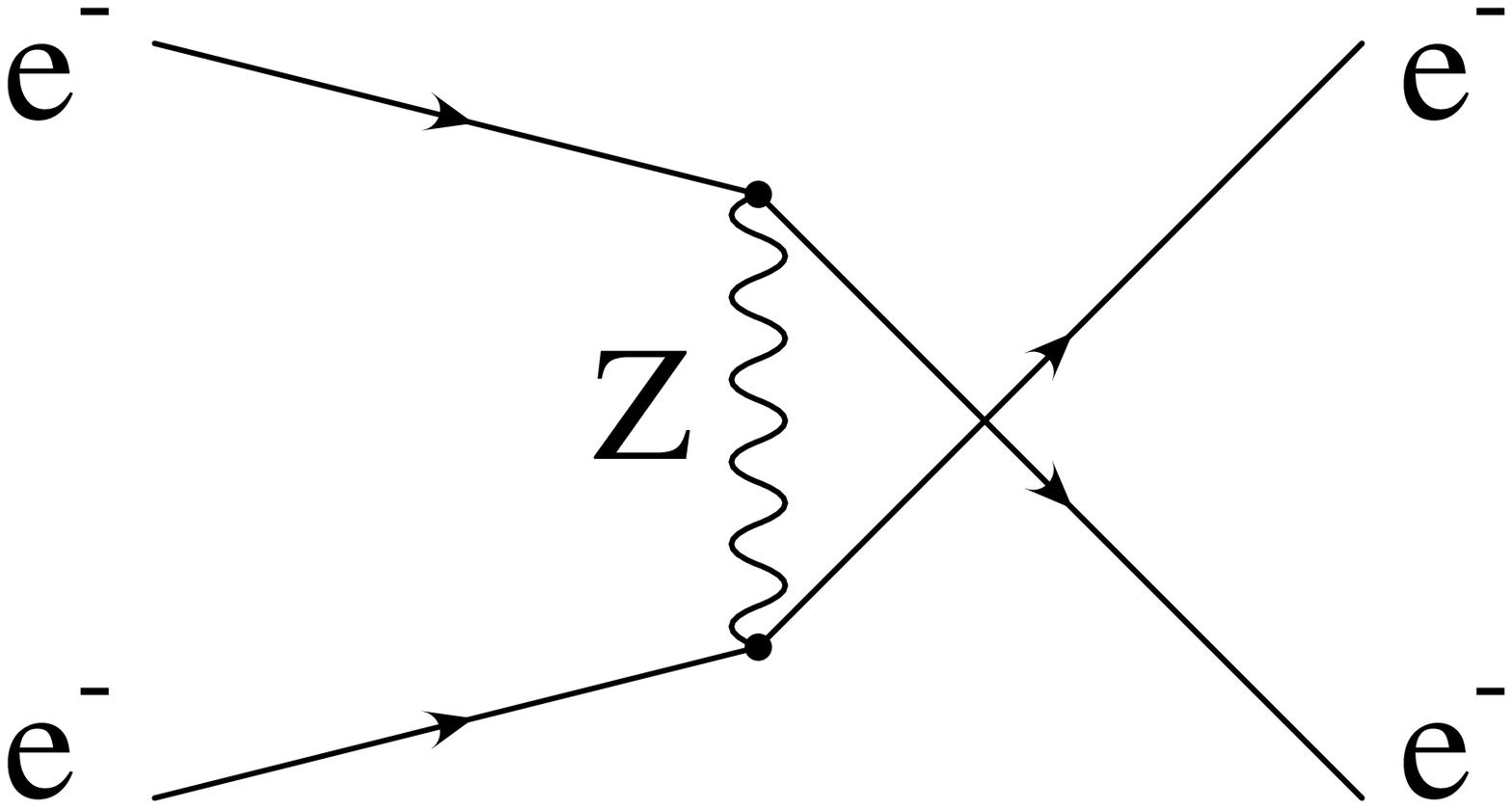,width=3.8cm,bbllx=210pt,bblly=410pt,%
bburx=630pt,bbury=550pt} 
\end{tabular}
}
$
\end{minipage}

\vspace*{15mm}
\noindent 
\fcaption{Neutral current direct and crossed $e^-e^-$ scattering 
amplitudes leading to the asymmetry $A_{LR}$ at tree level.}
\end{figure}
To simplify our discussion, we consider for illustration the
case $ys$ and $(1-y)s\gg m_Z^2$; so, terms of relative order $m_Z^2/ys
$ and  $m_Z^2/(1-y)s$ can be neglected. In that limit, one finds at
tree level
\be
{\dsll\over {\rm d}y} &=& \sigma_0 {1\over y^2(1-y)^2} {1\over
16\swq},
\nonumber \\
{\dsrr \over {\rm d}y} &=&\sigma_0 {1\over y^2(1-y)^2},
\nonumber \\
{\dslr\over {\rm d}y} &=& {\dsrl\over {\rm d}y}= \sigma_0 {y^4+(1-y)^4
\over y^2(1-y)^2} 
{1\over 4},
\label{eq:21}
\ee
and the asymmetries become
\be 
\alr^{(1)}(y) &=& {(1-4\ssw)(1+4\ssw)
\over 
1+16\sswq + 8\left[y^4+(1-y)^4\right]\sswq},
\label{eq:22} \\
\alr^{(2)}(y) &=& {(1-4\ssw)(1+4\ssw)
\over 
1+16\sswq},
\label{eq:23} 
\ee
Expanding about $\sw=1/4$, Eq.~(\ref{eq:23}) becomes
\be
\alr^{(2)}(y) = (1-4\sw)+{\cal O}[(1-4\sw)^2].
\ee
For arbitrary $s$ the asymmetries are maximal at $y=1/2$. At this
point we find, up to terms ${\cal O}[(1-4\sw)^2]$,
\be
\alr^{(1)}(y=1/2) &\approx & (1-4\sw)
{\frac{16\,x\,\left( 3 + 2\,x \right) }{3\,\left( 27 + 34\,x +
11\,{x^2} \right) }}, 
\nonumber \\[2mm]
\alr^{(2)}(y=1/2) &\approx & (1-4\sw){\frac{2\,x}{3 + 2\,x}},
\qquad x\equiv {s\over m_Z^2}.
\ee
General expressions for the asymmetries at any $s$ and $y$ are given
in the Appendix.

Because of the $(1-4\sw)$ dependence of
$\alr(e^-e^-)$, even with relatively modest angular coverage limited
to $0.1\le y\le 0.9$, \Mol scattering can be used to measure $\sw$
rather precisely, to about $\pm 0.0003$ at $\sqrt{s}\approx 1$
TeV. Although not likely to compete with potential very high
statistics $Z$ pole measurements, \Mol scattering can be used as a
powerful probe for ``new physics'' effects.  Indeed, for electron
composite effects parametrized by the four fermion interaction in
Eq.~(\ref{eq:fourf}), one finds $\Delta\alr \approx
sy(1-y)c_W^2/\alpha\Lambda^2$ for $e^-e^-$ \Mol scattering.  It can,
therefore, be more sensitive than $e^+e^-\to e^+e^-$ Bhabha
scattering\cite{BarklowH} (about 50\% better) and could probe $\Lambda
\sim 150$ TeV. 
That capability translates into a sensitivity to $m_{Z_\chi}\approx
12$ TeV for the additional neutral gauge boson, $Z_\chi$, of SO(10).
If there is a heavy doubly charged Higgs scalar $\Delta^{--}$ with
mass $m_\Delta$ which couples to $e^-_Le^-_L$ or $e^-_Re^-_R$ with
coupling $g_\Delta$, \Mol scattering would probe $g_\Delta^2/m_\Delta^2
\sim 5\times 10^{-5} G_F$, about four orders of magnitude improvement
beyond current bounds.  For $g_\Delta$ of order a typical gauge
coupling, that corresponds to $m_\Delta\gsim $ 10--20 TeV.

If one is interested in a more precise determination of $\sw$ via \Mol
scattering, extremely forward events must be detected. For example,
assuming detector acceptance down to about $5^\circ$ ($y=0.0019$),
Cuypers and Gambino\cite{Cuypers:1996it} have shown that $\Delta\sw
\approx \pm 0.0001$ may be possible at a $\sqrt{s}=2$ TeV $e^-e^-$
collider with $P_1=P_2=90\%$.

\begin{figure}
\noindent
\begin{minipage}{16.cm}
\hspace*{14mm}
$
\mbox{
\begin{tabular}{cc}
\psfig{figure=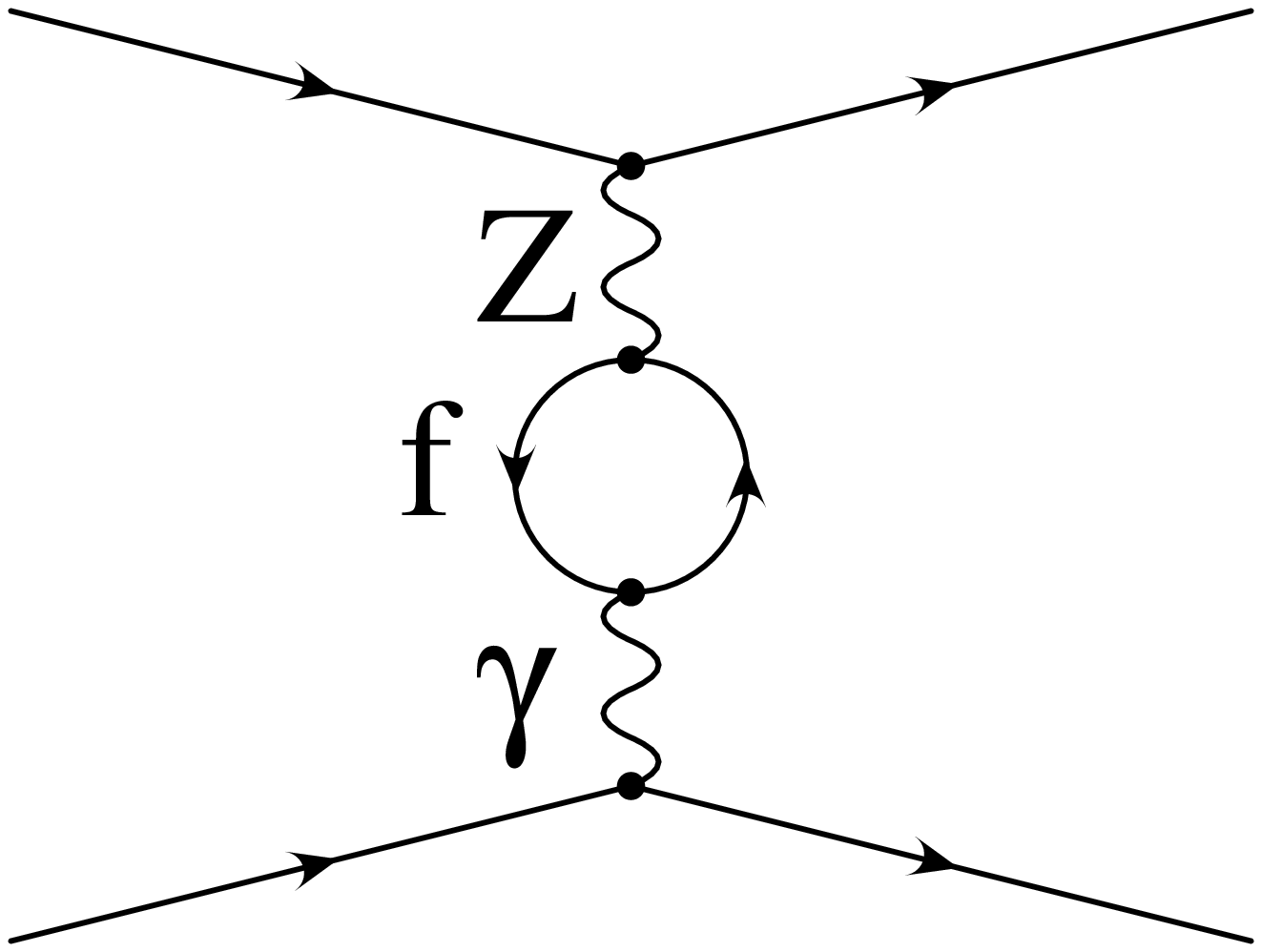,width=4.2cm,bbllx=210pt,bblly=410pt,%
bburx=630pt,bbury=550pt} &\hspace*{15mm}
\psfig{figure=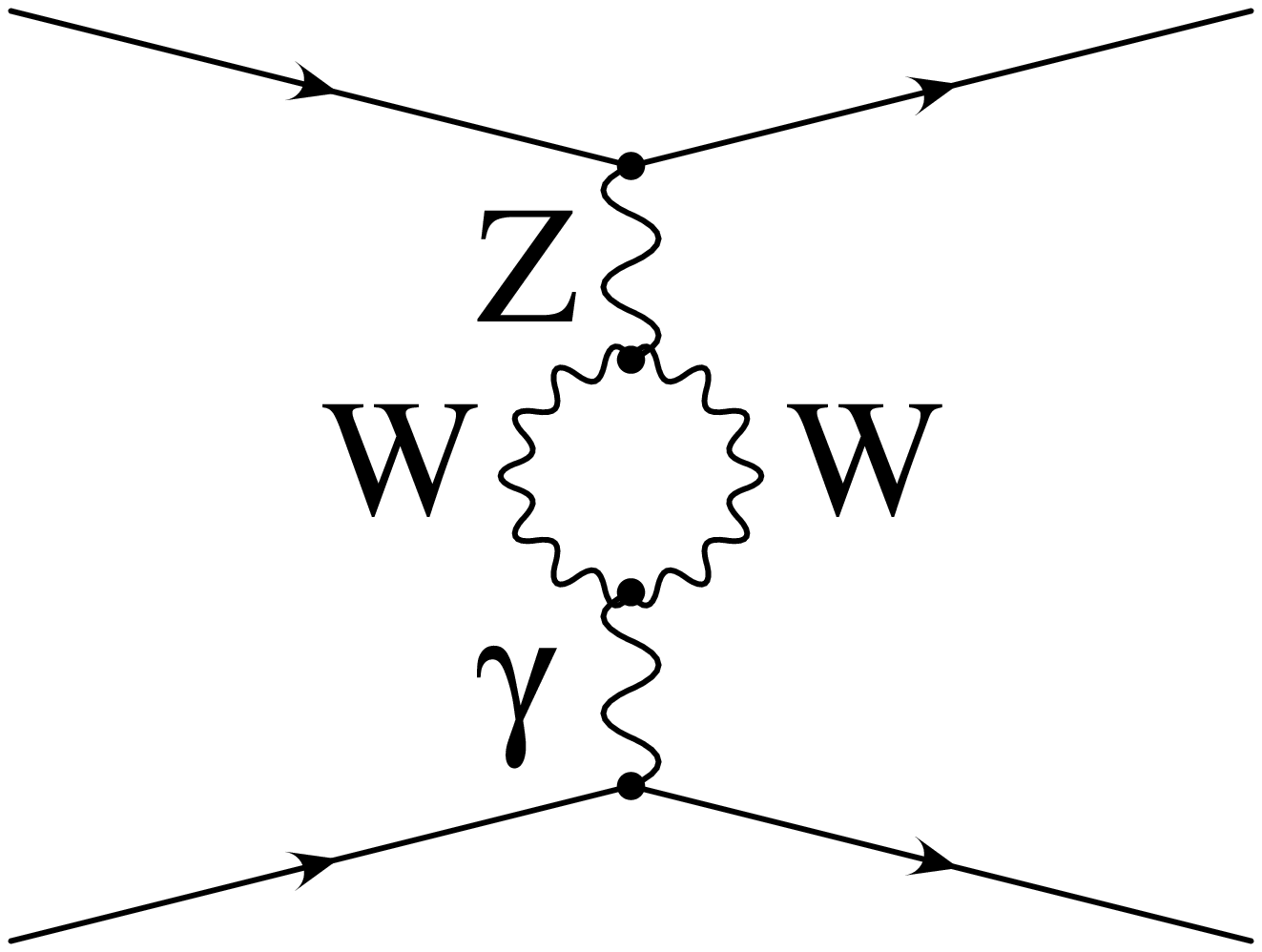,width=4.2cm,bbllx=210pt,bblly=410pt,%
bburx=630pt,bbury=550pt}
\\[20mm]
\psfig{figure=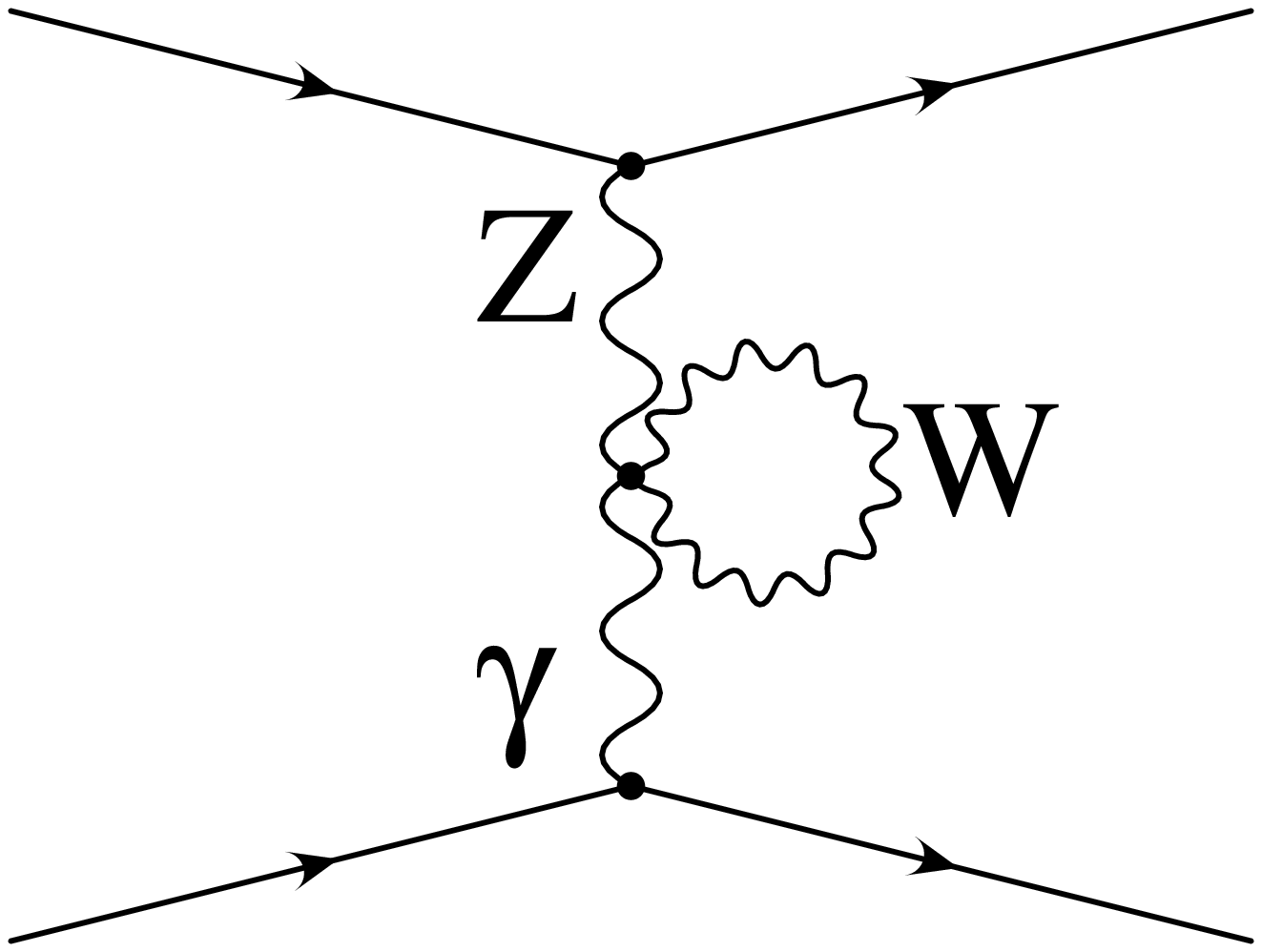,width=4.2cm,bbllx=210pt,bblly=410pt,%
bburx=630pt,bbury=550pt} &\hspace*{15mm}
\psfig{figure=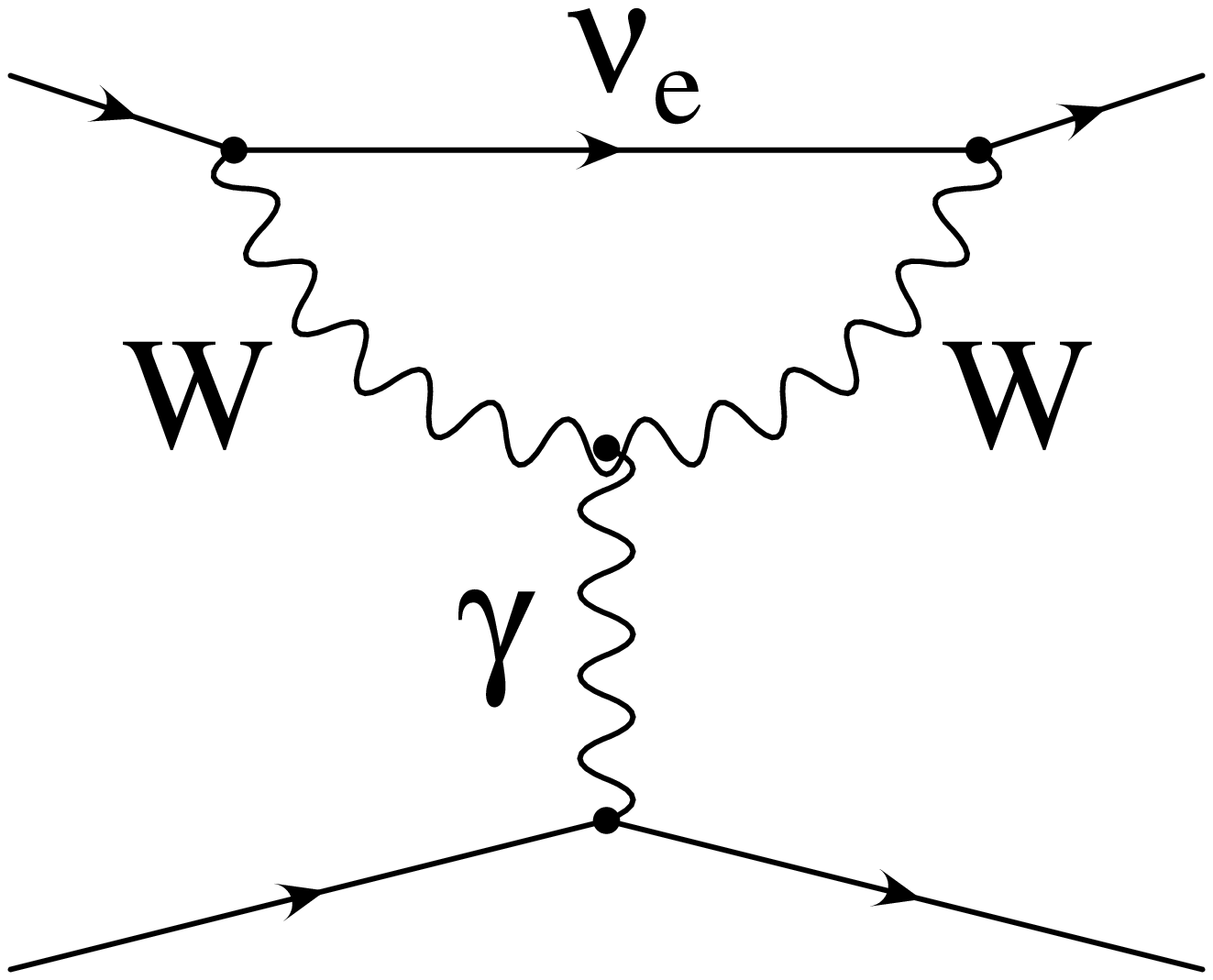,width=4.2cm,bbllx=210pt,bblly=410pt,%
bburx=630pt,bbury=550pt}\\[12mm]
\end{tabular}
}
$
\end{minipage}

\vspace*{4mm}
\noindent 
\fcaption{  $\gamma-Z$ mixing diagrams and  $W$-loop contribution to
the anapole moment.}
\end{figure}

For really precise studies of asymmetries, it will be necessary to
account for electroweak loop corrections. Such effects were examined
in detail for low $Q^2$ \Mol scattering and found to be very
significant\cite{Czarnecki:1996fw}. For $Q^2\approx 0.02$ GeV$^2$, of
relevance for an approved SLAC fixed target
experiment\cite{Kumar:1995ym}, loop corrections were shown to reduce
$\alr^{(1)}$ by about 40\%. That significant shift came mainly from
the running of $\sw(q^2)$\cite{Marciano:1981be}. It increases by about
3\% in going from $Q^2\approx m_Z^2$ to $\approx 0$. Because $\alr
^{(1)}$ is proportional to $1-4\sw$, that shift in $\sw$ gets
amplified in $\alr^{(1)}$.

Although we have not carried out a complete study of electroweak
radiative corrections to $\alr^{(1)}$ or $\alr^{(2)}$ for large $Q^2$,
we have examined the effect of a running $\sw(q^2)$. That running
incorporates the $\gamma Z$ mixing and charge radius diagrams
illustrated in fig.~2.  They effectively replace the tree level $\sw$
by $\kappa(q^2)\sw(m_Z)_\msbar$, where the current world average is 
\be
\sw(m_Z)_\msbar=0.2312 \pm 0.0002.
\ee
The radiative corrections in $\kappa(q^2)$ are given by (for
$Q^2=-q^2$)
\be
\kappa(Q^2) = \kappa_f(Q^2) + \kappa_b(Q^2)
\ee
where the subscripts denote $f=$ fermion loops and $b=$ boson
loops. Those quantities are found to be:
\be
\kappa_f(Q^2) &=&{1\over 3} \sum_f \left( T_{3f} Q_f - 2\sw^2
Q_f^2\right) 
\nonumber \\ &&\qquad\times
\left[
\ln {m_f^2\over m_Z^2} 
-{5\over 3} 
+4z_f
+(1-2z_f)p_f\ln {p_f+1\over p_f-1}
\right],
\nonumber \\
z_f &\equiv& {m_f^2\over Q^2}, \qquad p_f \equiv \sqrt{1+4z_f},
\ee
with $T_{3f}=\pm 1/2$, $Q_f=$ fermion charge, and the sum is over all
fermions. 
\be
\kappa_b(Q^2) &=& 
 - {42\cw+1\over 12}\ln \cw
 + {1\over 18}
\nonumber \\
&&
 -\left({p\over 2}\ln{p+1\over p-1}-1\right)
\left[
(7-4z)\cw +{1\over 6}(1+4z)
\right]
\nonumber \\
&&
 -z \left[
{3\over 4} - z 
+\left(z-{3\over 2}\right)p \ln {p+1\over p-1} 
+z \left( 2-z\right) \ln^2 {p+1\over p-1}
\right],
\nonumber \\
z &\equiv & {m_W^2\over Q^2}, \qquad p \equiv  \sqrt{1+4z}.
\ee
In fig.~3, we illustrate the dependence of $\sw(Q^2)$ on $Q^2$.  Note
that the relevant $Q^2$ for a given $\theta$ event are $ys$ and
$(1-y)s$ for direct and crossed diagrams in \Mol
scattering. Therefore, the running of $\sw$ should be incorporated at
the amplitude level. In that way, the final $\alr(y)$ will depend on
both $ys$ and $(1-y)s$ through $\sw$. 

\begin{figure}[htb]
\noindent
\begin{minipage}{16.cm}
\hspace*{15mm}
$
\mbox{
\hspace*{22mm}
\psfig{figure=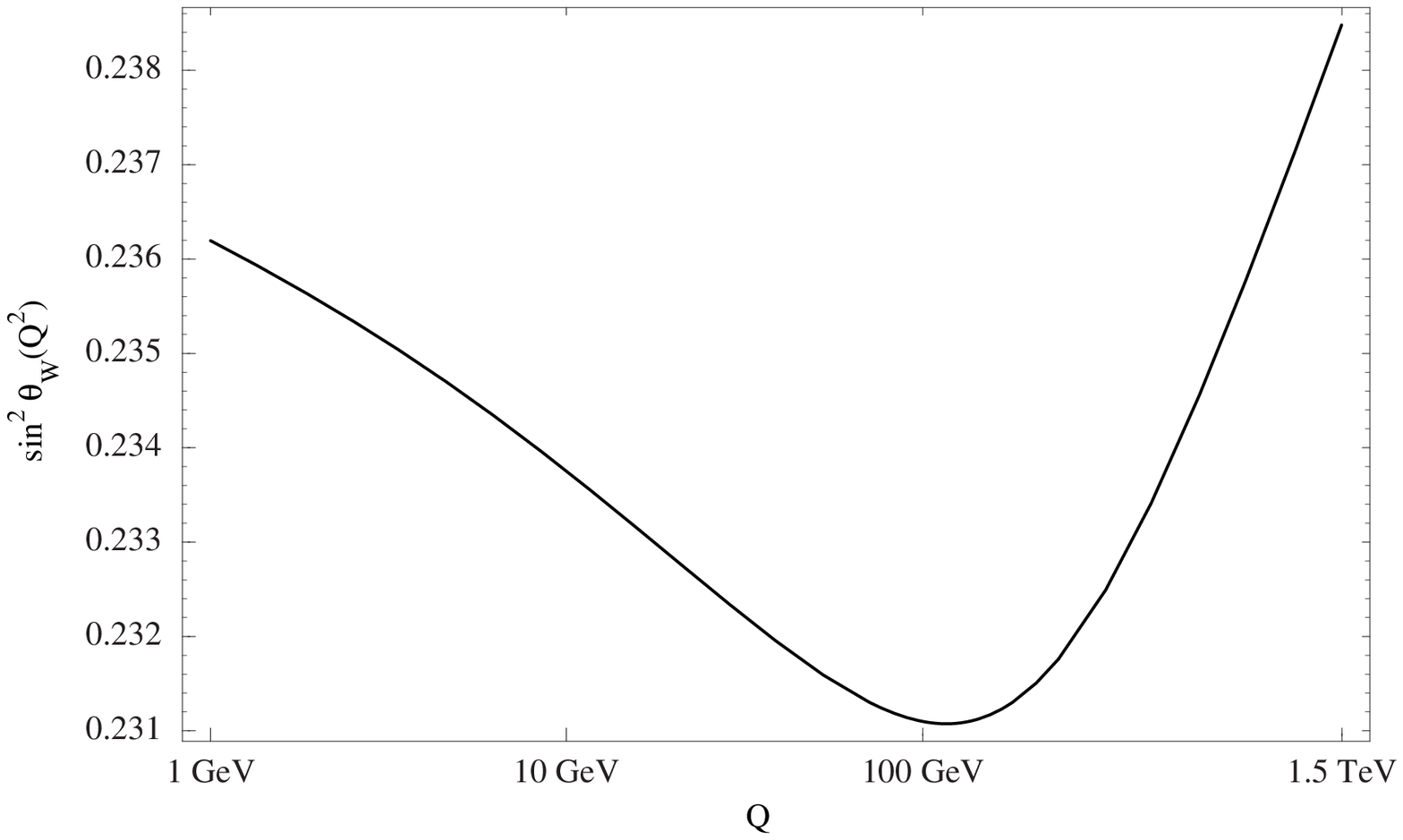,width=5.8cm,bbllx=210pt,bblly=300pt,%
      bburx=430pt,bbury=550pt} 
}
$
\end{minipage}

\vspace*{20mm}
\noindent 
\fcaption{Dependence of  $\sw(Q^2)$ on $Q$.}
\end{figure}

Note also that the variation in $\sw$ could be significant,
particularly at large $Q^2$. For studies of $\alr^{(2)}$ to even
several percent, the complete one loop calculation should be
included. Our calculation also demonstrates how a measurement of
$\alr^{(2)}(y)$ as a function of $y$ can be used (due to its $\sw$
sensitivity) to trace out the running of $\sw(Q^2)$.  The change in
$\sw(Q^2)$ by 3\% in going from $Q\approx m_Z$ to 1 TeV  translates
into about a 40\% reduction in $\alr^{(2)}$.

Parity violating left-right asymmetries have played key roles in
establishing the validity of the standard model.  From the classic
SLAC polarized $eD$ measurement to the $Z$ pole asymmetry, polarized
electron beams have proved their worth. They will continue to provide
valuable tools during the ILC era both in the $e^+e^-$ and $e^-e^-$
modes.  In the case of precision studies of parity violating
left-right scattering asymmetries, short--distance physics up to
${\cal O}$(150 TeV) will be indirectly explored. Even more exciting is
the possible direct detection of new phenomena such as supersymmetry
at these high energy facilities.  If ``new physics'' is uncovered,
polarization will help sort out its properties and decipher its place
in nature.

\nonumsection{Acknowledgements}
This work was supported  by the DOE under grant number
DE-AC02-76CH00016.

\nonumsection{References} 

\nonumsection{Appendix}

\begin{figure}[htb]
\noindent
\begin{minipage}{16.cm}
\hspace*{15mm}
$
\mbox{
\hspace*{24mm}
\psfig{figure=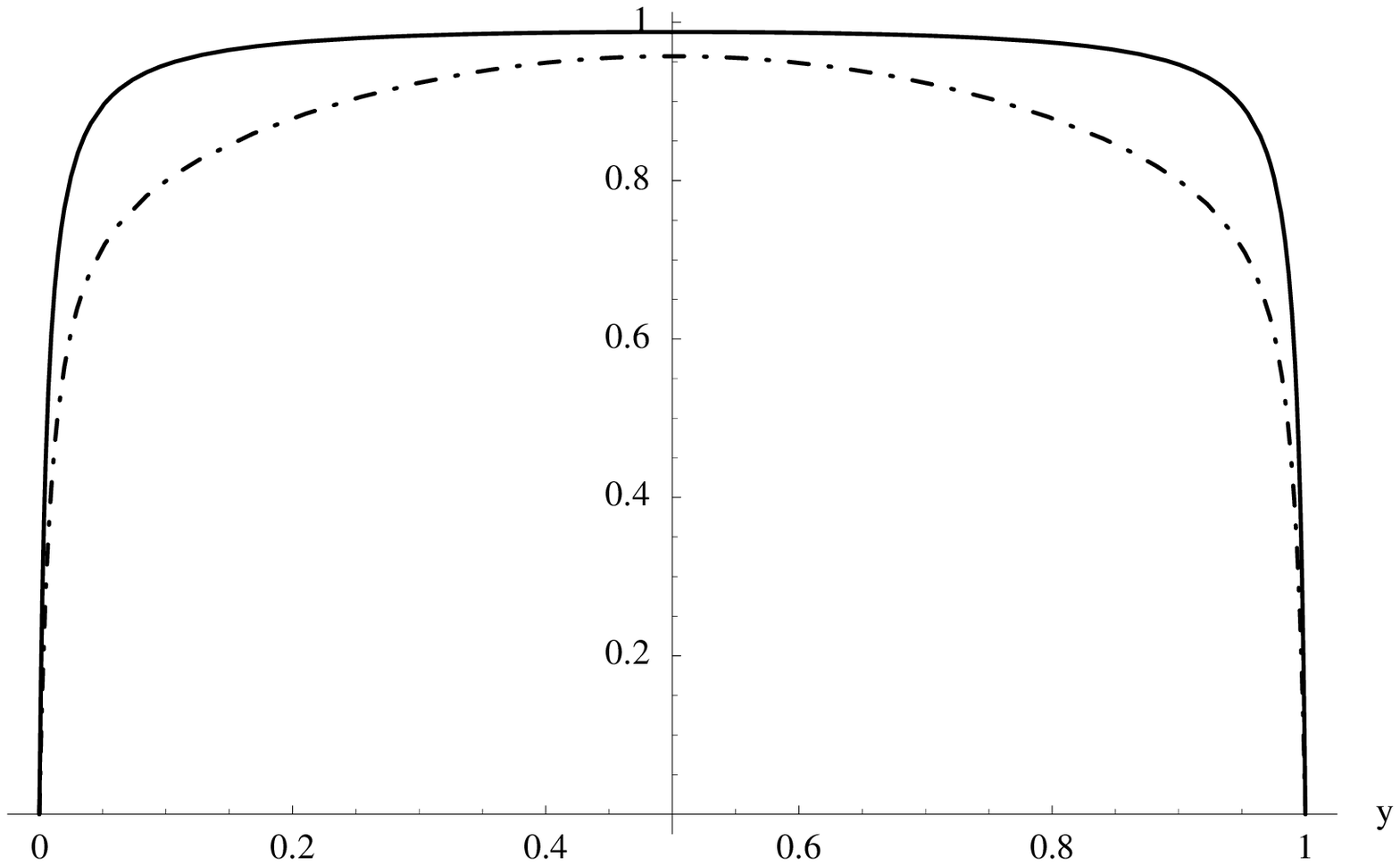,width=4.9cm,bbllx=210pt,bblly=300pt,%
      bburx=430pt,bbury=550pt} 
}
$
\end{minipage}

\vspace*{20mm}
\noindent 
\fcaption{Asymmetries $\alr^{(1)}$ (dashed line) and $\alr^{(2)}$
(solid) with the factor $1-4\sw$ divided out, as
functions of $y$.  We use $s=1$ TeV$^2$ and $\sw=1/4$.} 
\end{figure}

In this Appendix we present the formulas for the \Mol scattering
asymmetries for arbitrary $s$.  For the two asymmetries defined in 
Eqs.~(\ref{eq:16}) and (\ref{eq:17}) we find, up to terms ${\cal
O}[(1-4\sw)^2]$, 
\be
\alr^{(1)}(y) &\approx & (1-4\sw){4x(2+x)v[ 3(1+x)+2x(1+2x)v]
\over D},
\nonumber \\[2mm]
D&=& 18\,{{\left( 1 - v \right) }^2} 
 + 6\,\left( 6 - 11\,v + 9\,{v^2} \right) \,x + 
  \left( 18 + 6\,v - 7\,{v^2} + 22\,{v^3} \right) \,{x^2} 
\nonumber \\&& 
+ 
  36\,v\,\left( 1 - v + {v^2} \right) \,{x^3} + 
  4\,{v^2}\,\left( 5 - 4\,v + 2\,{v^2} \right) \,{x^4},
\nonumber \\[2mm]
\alr^{(2)}(y) &\approx & (1-4\sw)
{4x(2+x)v\over 3(1+x)+2x(1+2x)v},
\ee
where $ x\equiv {s\over m_Z^2}$ and $v \equiv y(1-y)$.

\end{document}